\documentclass[%
 reprint,
 amsmath,amssymb,
 aps,
]{revtex4-1}

\usepackage{color}
\usepackage{graphicx}
\usepackage{dcolumn}
\usepackage{bm}

\begin{document}

\preprint{APS/123-QED}

\title{Reconciling Nuclear and Astrophysical Constraints}

\author{V. Dexheimer}
\affiliation{Department of Physics, Kent State University, Kent OH 44242 USA}
\email{vdexheim@kent.edu}

\author{R. Negreiros}
\affiliation{Instituto de Fisica, Universidade Federal Fluminense, Niteroi, Brazil}

\author{S. Schramm}
\affiliation{FIAS, Johann Wolfgang Goethe University, Frankfurt, DE}

\date{\today}

\begin{abstract}
In view of new constraints put forth by recent observations and measurements in the realm of astrophysics and nuclear physics, we update the non-linear realization of the sigma model as to reflect such constraints. By doing this, we obtain new equations of state that may be used to describe neutron stars. Such equations of state are obtained by investigating different ways by which the vector mesons self-interact. Furthermore, we also investigate the role played by the delta mesons in the model. As a result, we are able to develop equations of state that are in better agreement with data, such as nuclear compressibility and slope of the symmetry energy at saturation, star masses, radii, and cooling profiles.
\end{abstract}

\pacs{Valid PACS appear here}
\maketitle


In recent years, measurements of massive neutron stars with masses around $2.0$ M$\odot$ \cite{Demorest:2010bx,Antoniadis:2013pzd} have been used to select equations of state that fulfill astrophysics constraints. In addition, estimates of neutron star radii, which seem to be smaller than previously thought, have become much more sophisticated \cite{Psaltis:2013fha,Guillot:2013wu,Lattimer:2013hma,Heinke:2014xaa}. These have been used with the same purpose. 
Note that, in addition to more direct radii measurements, trustworthy theoretical calculations can provide insight into this quantity \cite{Hebeler:2010jx,Horowitz:2001ya,Chen:2014sca,Ozel:2015fia,Chen:2015zpa}. Still, until more accurate radii measurements are made, it is important to explore a somewhat large range of values.

Nuclear physics constraints have also improved, and quantities such as the symmetry energy and its slope have had their values narrowed at saturation density to $29.0$ MeV $<E_{\rm{sym}}< 32.7$ MeV and $40.5$ MeV $<L< 61.9$ MeV in the analysis of Ref. \cite{Lattimer:2012xj}, which we take as a guideline for the underlying investigation. Other works have also found similar values \cite{Li:2013ola}. For a recent review on the symmetry energy, please see Ref.~\cite{Horowitz:2014bja}.

In view of the aforementioned improvements in the analysis of astrophysical and nuclear data, specifically the slope of the symmetry energy at saturation density, we present an updated version of the hadronic chiral SU(3) sigma model. Following the idea proposed recently in Refs.~\cite{Steiner:2012rk,Dutra:2014qga,Bizarro:2015wxa}, we study different possible self-interactions of the vector mesons. We do this by considering different possible coupling schemes, but restricting ourselves to  only those that are chirally invariant.

We also investigate the influence of the delta meson on the model and on its predicted microscopic quantities and macroscopic observables. We recall that effects of the delta meson in relativistic mean field models have been investigated in Refs.\cite{Kubis:1997ew,Liu:2001iz,Gomes:2014aka,Menezes:2004vr} and many others.

Chiral sigma models are effective relativistic models that describe hadrons interacting via meson exchange and, most importantly, are constructed from symmetry relations. They are constructed in a chirally invariant manner since the particle masses originate from interactions with the medium and, therefore, go to zero at high density and/or temperature. The underlying extended sigma model is in very good agreement with nuclear physics data, such as the vacuum masses of the baryons, saturation properties, hyperon potentials, and pion and kaon decay constants $f_\pi$ and $f_\kappa$, etc. \cite{Papazoglou:1998vr,Bonanno:2008tt}. 

The Lagrangian density of the SU(3) non-linear sigma model, called hereafter the chiral mean-field (CMF) model, applied to neutron star matter can be found in Refs.~\cite{Dexheimer:2008ax,Schurhoff:2010ph}. A recent extension of this model also includes quarks as dynamical degrees of freedom \cite{Dexheimer:2009hi,Negreiros:2010hk,Steinheimer:2010ib,2013PhRvC..88a4906H,Dexheimer:2014pea}. In this work, we make use of a simple hadronic version of the model without hyperons or quarks. The reason behind such a choice is that we want to study the effect of different possible self-interactions of the vector mesons, without the interference of the appearance of hyperons or quarks in the star. The CMF Lagrangian density of the model we use in this work reads
\begin{eqnarray}
\mathcal{L} = \mathcal{L}_{\rm{Kin}}+\mathcal{L}_{\rm{Int}}+\mathcal{L}_{\rm{Self}}+\mathcal{L}_{\rm{SB}}\,,
\end{eqnarray}
where, besides the kinetic energy term for hadrons and leptons (included to ensure charge neutrality), the terms:
\begin{eqnarray}
\mathcal{L}_{\rm{Int}}&=&-\sum_i \bar{\psi_i}[\gamma_0(g_{i\omega}\omega+g_{i\phi}\phi+g_{i\rho}\tau_3\rho)+M_i^*]\psi_i,\nonumber\\
\mathcal{L}_{\rm{Self}}&=&+\frac{1}{2}(m_\omega^2\omega^2+m_\rho^2\rho^2+m_\phi^2\phi^2)\nonumber\\
&-&k_0(\sigma^2+\zeta^2+\delta^2)-k_1(\sigma^2+\zeta^2+\delta^2)^2\nonumber\\&-&k_2\left(\frac{\sigma^4}{2}+\frac{\delta^4}{2}
+3\sigma^2\delta^2+\zeta^4\right)
-k_3(\sigma^2-\delta^2)\zeta\nonumber\\&-&k_4\ \ \ln{\frac{(\sigma^2-\delta^2)\zeta}{\sigma_0^2\zeta_0}}+L_{\rm{vec} 4}\,,\nonumber\\
\mathcal{L}_{\rm{SB}}&=&-m_\pi^2 f_\pi\sigma-\left(\sqrt{2}m_k^ 2f_k-\frac{1}{\sqrt{2}}m_\pi^ 2 f_\pi\right)\zeta\,,
\end{eqnarray}
represent the interactions between baryons and vector and scalar mesons, the self interactions of scalar and vector mesons, and an explicit chiral symmetry breaking term, which is responsible for producing the masses of the pseudo-scalar mesons. The index $i$ denotes the protons and neutrons. The electrons are included as a free Fermi gas. The mesons included are the vector-isoscalars $\omega$ and $\phi$ (strange quark-antiquark state), the vector-isovector $\rho$, the scalar-isoscalars $\sigma$ and $\zeta$ (strange quark-antiquark state) and  the scalar-isovector $\delta$, with $\tau_3$ being twice the isospin projection of each particle ($\pm$1). The isovector mesons affect isospin-asymmetric matter and thus, are important for neutron star physics. Also, the $\delta$ meson has a contrary but complementary role to the $\rho$ meson, much like the $\sigma$ and $\omega$  mesons.

Note that the strange mesons influence the model through the coupling to the nucleons ($\zeta$) and through self-interaction terms ($\phi$). The different possible self-interaction terms of the vector mesons that are chiral invariant (referred to as $L_{\rm{vec} 4}$ above) are the following
\begin{itemize}
\item $\mathcal{L}_{\rm{vec} 4} = 2 Tr(V^4)$\,,
\item $\mathcal{L}_{\rm{vec} 4} = [Tr(V^2)]^2$\,,
\item $\mathcal{L}_{\rm{vec} 4} = [Tr(V )]^4/4$\,,
\end{itemize}
where $V$ stands for the matrix of the vector meson multiplet, which reduces to a diagonal form in the mean field limit, i. e.,
\begin{eqnarray}
V = \left( \begin{array}{ccc}
(\omega + \rho)/\sqrt{2} & 0 & 0 \\
0 & (\omega - \rho)/\sqrt{2} & 0 \\
0 & 0 & \phi \end{array} \right) \,.
\end{eqnarray}
The different self-interaction terms of the vector mesons shown above correspond, respectively, to the coupling schemes C1, C3, and C4 in the following. We also introduce the coupling scheme C2, which is a linear combination of 1 and 3, and presents no $\omega \rho$ mixing. Note that the coupling scheme C4 is quite special, as it contains contributions that are linear in the isoscalar vector field
\begin{itemize}
\item C1: $\mathcal{L}_{\rm{vec} 4} = g_4 (\omega^4 + 6 \omega^2 \rho^2 + \rho^4 + 2 \phi^4)$\,,
\item C2: $\mathcal{L}_{\rm{vec} 4} = g_4 (\omega^4 + \rho^4 + \phi^4/2 + 3 \rho^2 \phi^2 + 3 \omega^2 \phi^2)$\,,
\item C3: $\mathcal{L}_{\rm{vec} 4} = g_4 (\omega^4 + 2 \omega^2 \rho^2 + \rho^4 + 2 \omega^2 \phi^2 + \phi^4 +2 \rho^2 \phi^2)$\,,
\item C4: $\mathcal{L}_{\rm{vec} 4} = g_4 (\omega^4 + 3 \omega^2 \phi^2 + \phi^4/4 + 4 \omega^3 \phi/\sqrt{2} + 2 \omega \phi^3/\sqrt{2})$\,.
\end{itemize}

The effective masses for the baryons are simply generated by the scalar mesons, except for a small explicit mass term $M_0$ 
\begin{eqnarray}
M_{i}^*&=&g_{i\sigma}\sigma+g_{i\delta}\tau_3\delta+g_{i\zeta}\zeta+M_{0_i}\,.\end{eqnarray}

The coupling constants of the model were modified from Ref.~\cite{Dexheimer:2008ax}, in the sense that we used ranges, rather than values, for nuclear constraints at saturation, such as baryon density ($\rho_0=0.15$ to $0.17$ fm$^{-3}$), binding energy ($B/A=-15.75$ to $-16.3$ MeV), and symmetry energy ($E_{\rm{sym}}=29 - 32.7$ MeV). Within this setup, we try to achieve the minimum possible compressibility $K$ and symmetry energy slope $L$, while still fulfilling astrophysical constraints.  For a review on compressibility values see Ref. \cite{Stone:2014wza} and references therein.

\begin{figure}[t]
\vspace{3mm}
\includegraphics[width=9.5cm]{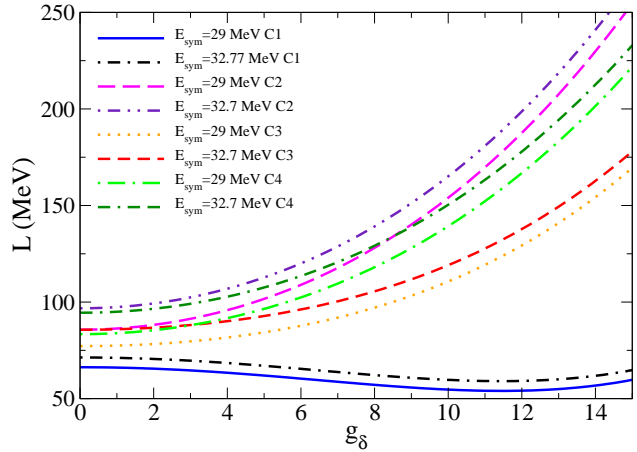}
\caption{(Color online) Slope of the symmetry energy at saturation as a function of the coupling constant of the $\delta$ meson for different coupling schemes.\label{L}}
\end{figure}

More specifically, we vary the vector sector of the model, $g_4$, $g_{\omega}$ and $g_{\rho}$ ($g_{\phi}=0$) while keeping most of the the scalar sector constant ($g_{\sigma}=-9.83$, $g_{\zeta}=1.22$, $k_0=1.19 \chi^2$, $k_1=-1.40$, $ k_2=5.55$, $k_3=2.65 \chi$ and $k_4=-0.06 \chi^4$, with $\chi = 401.93$ MeV). The scalar sector of the model cannot be further adjusted as it was fixed to reproduce the vacuum masses of the baryons and scalar mesons, and the pion and kaon decay constants $f_\pi$ and $f_\kappa$. Only $g_{\delta}$ is modified freely, as it only affects isospin asymmetric matter.

We now present microscopic and macroscopic properties reproduced by the 4 coupling schemes C1 to C4. We fit the coupling constants $g_4$ and $g_{\omega}$ to reproduce nuclear and astrophysical constraints. We set $g_{\rho}$ so as to obtain two curves, one for the lowest accepted value from the  range of the symmetry energy (29 MeV) and, another for the corresponding highest value (32.7 MeV). Finally, we study the behavior of different quantities as a function of the strength of the delta meson coupling to the nucleons $g_\delta$. 

\begin{figure}[t]
\vspace{3mm}
\includegraphics[width=9.5cm]{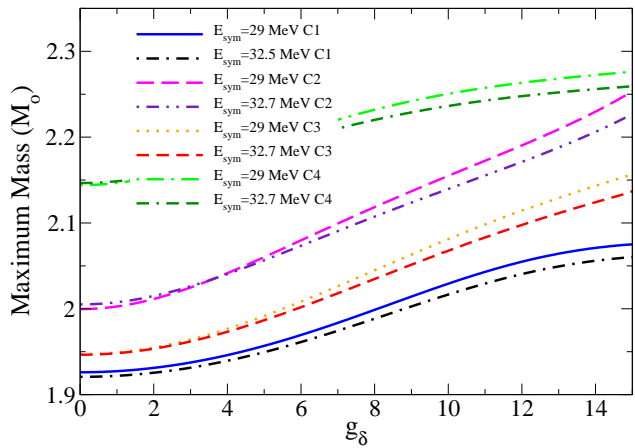}
\caption{(Color online) Maximum possible star mass as a function of the coupling constant of the $\delta$ meson for different coupling schemes.\label{Mass}}
\end{figure}

Within this setup, it is interesting to see the behavior of $L$ (the slope of the symmetry energy at saturation) as we vary the delta coupling. As shown in Fig.~\ref{L},  although the value of $L$ increases with the delta coupling for cases C2, C3, and C4, it reaches a minimum value for case C1 around $g_\delta=11.5$. This is related to the strength of the $\omega\rho$ term in the vector-meson self-interactions (present in C1 and C3 cases), which induces a softer equation of state for asymmetric matter at higher densities (due to lower absolute values of $\omega$ and $\rho$ necessary to fit nuclear saturation and asymmetry properties). Note that, even when $g_\delta=11.50$, the coupling strength of the delta meson is still small compared with that of the sigma meson $(g_{\delta}/m_\delta)^2/(g_{\sigma}/m_\sigma)^2\sim0.3$. As expected, $L$ is higher for higher values of the symmetry energy for all coupling schemes, as pointed out in Refs.~\cite{Oyamatsu:2002mv,Gandolfi:2011xu,Tsang:2012se,Lopes:2014wda}, where a linear behavior between $L$ and $E_{\rm{sym}}$ was found.

The respective maximum masses achieved for stars generated by each coupling scheme are calculated through the use of the Tolman--Oppenheimer--Volkoff (TOV) equations and are shown in Fig.~\ref{Mass}. Once more, the strength of the $\omega\rho$ coupling plays an important role, as it renders the equation of state softer, which in turn yields less massive stars. Nevertheless, the maximum masses for all the coupling schemes analyzed increase with $g_\delta$, since this isovector channel generates a larger asymmetry between protons and neutrons (at low and intermediate densities) and, therefore, more Fermi pressure in the star. 

As already mentioned, the coupling scheme C4 for the self-interaction of the vector mesons does not follow the trend set by the other schemes, in the sense that it contains a term which is linear on the strange vector meson $\phi$. This feature modifies the behavior of the model, which has to be treated separately from the previous coupling schemes. For example, it has a bare mass for the nucleons of $M_0=50$ MeV. This allows for smaller compressibility (which otherwise would be large) in better agreement with nuclear physics data. Note that this version of the vector-meson self-interaction fitted to only one value of symmetry energy has already been discussed in Refs.~\cite{Dexheimer:2008ax,Schurhoff:2010ph,Dexheimer:2009hi,Negreiros:2010hk,2013PhRvC..88a4906H}.

\begin{figure}[t]
\vspace{3mm}
\includegraphics[width=9.5cm]{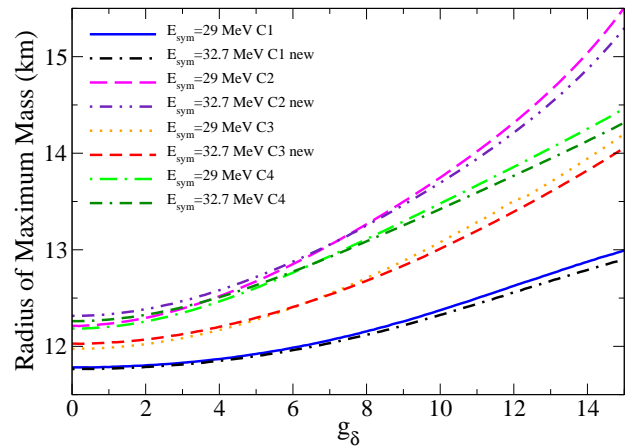}
\caption{(Color online) Radius of the maximum possible star mass as a function of the coupling constant of the $\delta$ meson for different coupling schemes.\label{radius}}
\end{figure}

In order to discuss the radii of stars, it is important to include a separate equation of state for the crust. In this work, we choose to use the Baym-Pethick-Sutherland (BPS) equation of state containing an inner crust, an outer crust and an atmosphere \cite{Baym:1971pw}. The respective radii for each maximum mass star (shown in Fig.~\ref{Mass} for each coupling scheme) are shown in Fig.~\ref{radius}. It is clear that the radii behave in the opposite manner with respect to the increase of $g_\delta$: the radius increases and is in worse agreement with data \cite{Psaltis:2013fha,Guillot:2013wu,Lattimer:2013hma,Heinke:2014xaa}. Note that, for the coupling scheme C4, neither the maximum mass nor the respective star radius increases much with $g_\delta$, when compared with the other coupling schemes. This is again related to change of the value of the meson fields (specifically the lack of change in the meson $\sigma$, the counterpart of $\omega$) with $g_\delta$, which affects the equation of state. If, instead of radii for maximum mass stars we had shown the radii for $1.4$ $M_\odot$ stars, Fig.~\ref{radius} would look similar, except that all radii would be shifted up by about $2$ km.

\begin{figure}[t]
\vspace{3mm}
\includegraphics[width=9.5cm]{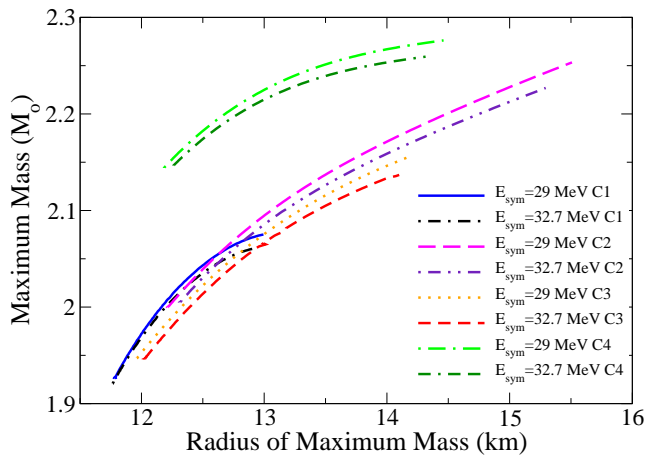}
\caption{(Color online) Maximum possible star masses and respective radii for each coupling schemes obtained by varying $g_\delta$ from $0$ to $15$.\label{MxR}}
\end{figure}

At this point, it becomes clear that the influence of  the $\delta$ meson on isospin asymmetric equations of state has to be carefully analyzed. This becomes even more evident when we look at Fig.~\ref{MxR}, which shows the increase of a star's possible maximum mass and respective radius in relation to the different coupling schemes. At first, it looks like there is no reconciliation between large mass and small radius stars. However, after careful analysis, we can identify a parameter set that fulfills all the constraints discussed. Such a parameter set is shown in Table~\ref{tabela1}.  Note, in particular, that the parametrization chosen for C1 reproduces small slopes for the symmetry energies (for the whole range of symmetry energies at saturation) and yet still yields  massive stars.

Finally, as a complementary approach to the usual study of neutron star macroscopic properties, we briefly discuss their cooling. The cooling or thermal evolution of neutron stars is strongly dependent on whether or not the direct Urca process (DU henceforth) takes place, as well as on pairing among the nucleons. Given efficient neutrino generation in the DU process, if it takes place in the star, it will lead to a fast (sometimes also called enhanced) thermal evolution of the object. This in turn, may lead to disagreement with observed thermal data on compact objects. The efficiency of the DU process highlights the need for an accurate description of pairing among nucleons in neutron stars, since it suppresses the neutrino emission from the DU process (among other processes as well) and may thus reconcile theory with observation. For a detailed review on the neutrino emission of neutron stars as well as of pairing see Ref. \cite{Yakovlev2001}.

\begin{figure}[t]
\vspace{3mm}
\includegraphics[width=9.5cm]{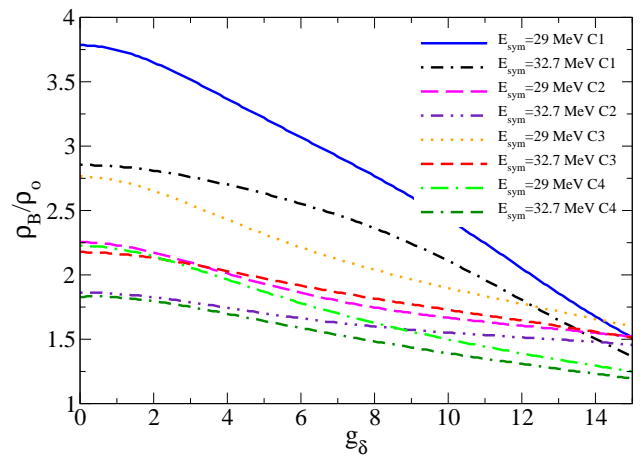}
\caption{(Color online) Normalized density threshold for the direct Urca process as a function of the coupling constant of the $\delta$ meson for different coupling schemes.\label{Urca}}
\end{figure}

Figure~\ref{Urca} shows the density at which the DU process conditions are achieved, thus signaling its onset. This density is much higher (and the process consequently only takes place much more towards the center of the star) for the coupling schemes C1 and C3. These are again the cases in which there is a strong $\omega\rho$ term for the vector-meson self-interaction. Also, the DU threshold density decreases quickly with the increase of $g_\delta$, as the star becomes more isospin symmetric (more protons at large densities) due to lower absolute values of the $\rho$ meson. Finally, note that larger symmetry energies are associated with lower density thresholds. This is only natural, since the symmetry energy is directly related to the energy balance between symmetric and neutron matter. In other words, larger symmetry energy means larger energy cost to create more neutrons and, consequently larger proton fraction. The relation between symmetry energy and neutron star cooling has been studied, for example, in Ref. \cite{Lattimer:1991ib,Steiner:2006bx,Cavagnoli:2011ft,2013ApJ...779L...4N}.

\begin{table*}
\caption{\label{tabela1} Best fit for the coupling constants of the model using different vector-meson self-interactions which fulfill nuclear physics and astrophysics constraints. The radius shown is the one corresponding to the star's maximum mass.}
\begin{ruledtabular}
\begin{tabular}{c|ccccc|ccccc}
$\ \ C\ \ $  & $g_{4}$&$g_{\omega}$ & $g_{\rho}$ & $g_{\delta}$ & $M_{0}$ (MeV) & $K$ (MeV) & $E_{\rm{sym}}$ (MeV) & $L$ (MeV) & $M$ ($M_\odot$) & $R$ (km) \\
\hline
$ C1$ &$58.40$&$13.66$&$11.06 $&$11.50 $&$0 $&$271.41$&$ 29.00$  &$ 54.00$  &$ 2.05 $&$ 12.56  $\\
$ C1 $&$58.40$&$13.66$&$11.33 $&$11.50 $&$0 $&$271.41$&$32.70$ &$ 59.05$  &$ 2.04  $&$ 12.49  $\\
$ C2 $&$58.40$&$13.66$&$3.51 $&$0 $&$0 $&$271.41$&$ 29.00$ &$ 85.68$  &$ 2.00  $&$ 12.22  $\\
$ C2 $&$58.40$&$13.66$&$4.02 $&$0 $&$0 $&$271.41$&$ 32.70$ &$ 96.77$  &$ 2.01  $&$ 12.32  $\\
$ C3 $&$58.40$&$13.66$&$3.82 $&$ 0 $&$0 $&$271.41$&$ 29.00$ &$ 77.19$  &$ 2.12  $&$ 14.38 $\\
$ C3 $&$58.40$&$13.66$&$4.38 $&$0 $&$0 $&$271.41$ &$ 32.70$  &$85.64$ &$ 1.95  $&$ 12.01 $\\
$ C4 $&$34.20$&$11.88$&$3.70 $&$ 0 $&$150$& $292.91$ &$ 29.00$ &$ 83.38$  &$ 2.14 $&$ 12.18  $\\
$ C4 $&$34.20$&$11.88$&$4.19$&$0 $&$150$ &$292.91$&$ 32.70$ &$ 94.48$  &$ 2.15 $&$ 12.26  $\\
\end{tabular}
\end{ruledtabular}
\end{table*}

\begin{figure}[t]
\vspace{3mm}
\includegraphics[width=8.9cm,clip,trim=5 0 0 0]{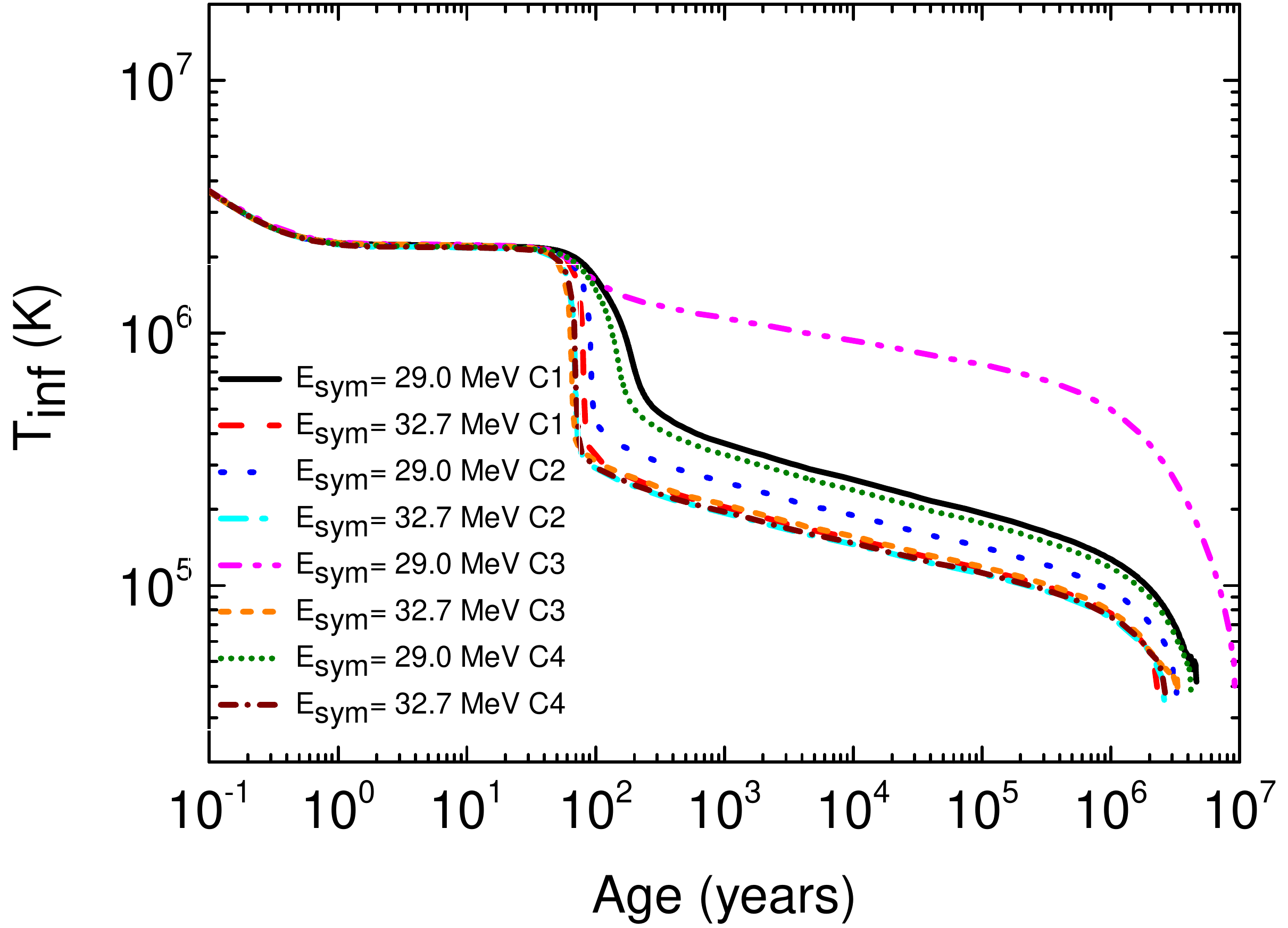}
\caption{(Color online) Surface temperature as observed at infinity of neutron stars with $1.4$ M$_\odot$ for the parameters described in Table~\ref{tabela1}. \label{cool_1.4} }
\end{figure}

\begin{figure}[t]
\vspace{3mm}
\includegraphics[width=8.9cm,trim=5 0 0 20]{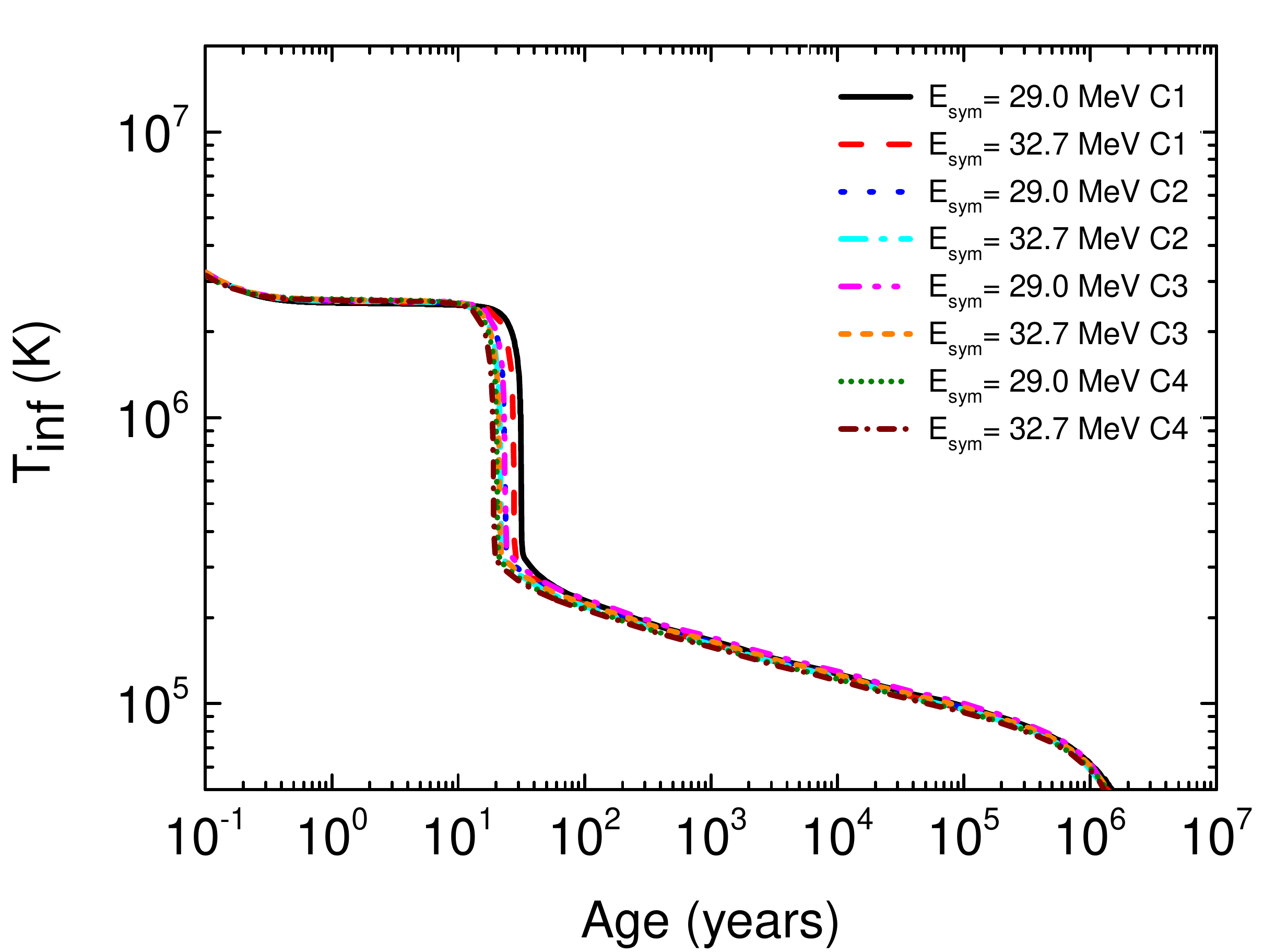}
\caption{(Color online) Same as Fig.~\ref{cool_1.4} but for neutron stars with the maximum possible mass. \label{cool_max}}
\end{figure}

To complement the previous discussion, we perform a preliminary study of the thermal evolution of neutron stars using the coupling schemes presented in this paper. More precisely, we performed full thermal evolution simulations of neutrons stars whose inner compositions are described by the parameter sets outlined in Table \ref{tabela1}. All relevant thermal process are consistently taken into account. At first, we do not consider any pairing in order to be able to clearly identify the effects of each coupling scheme on the thermal evolution. For this case, we show in Figs.~\ref{cool_1.4} and \ref{cool_max} the thermal evolution of stars with $M = 1.4\,M_\odot$ and the maximum mass of each respective parameter set.

Figures~\ref{cool_1.4} and \ref{cool_max} show the effects of the onset of the DU process on the thermal evolution of the stars. As can be seen in Fig.~\ref{cool_1.4}, the coupling scheme that exhibits the ``slowest" cooling is C3 with a symmetry energy of 29 MeV. This stems from the fact that for stars with that mass the DU process onset is not achieved within this parameter set (in agreement with Fig.~\ref{Urca}). As for stars of the maximum mass, the DU onset is achieved in every model, as shown in Fig.~\ref{cool_max}.

Finally, to illustrate the importance of pairing, we perform cooling simulations allowing the nucleons to form pairs. The neutrons are allowed to form singlet pairs in the crust and triplets in the core, whereas the protons are allowed to form singlet pairs in the core. A similar description has been used to analyze the thermal evolution of the neutron star in the supernova remnant Cassiopeia A \cite{Shternin2011,Page2011a}. We note that, in addition to pairing, we have also included pair breaking-formation (PBF) neutrino emission process.

\begin{figure}[t]
\vspace{3mm}
\includegraphics[width=8.9cm,trim=5 0 0 0]{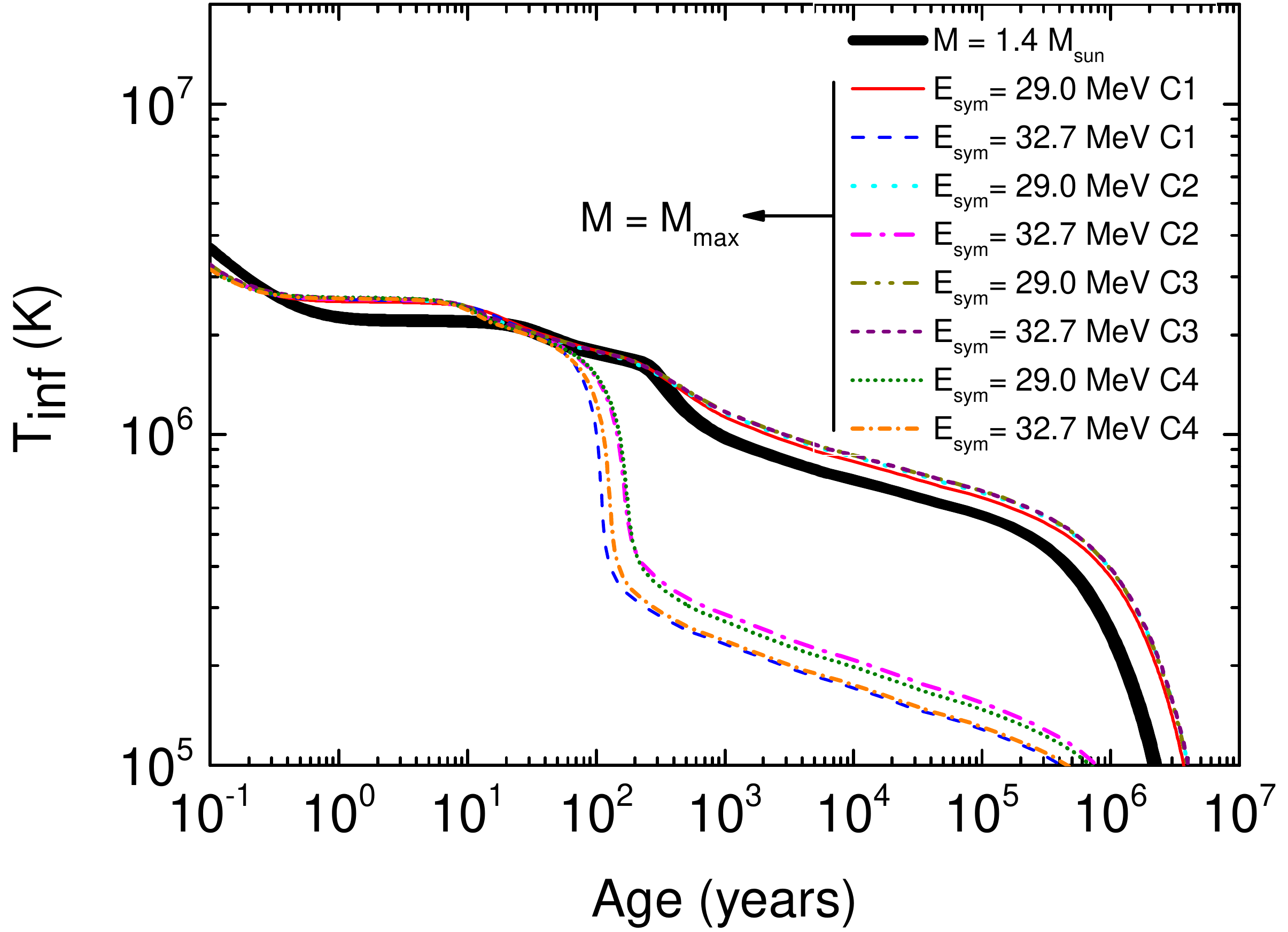}
\caption{(Color online) Combination of Figs~\ref{cool_1.4} and \ref{cool_max} with the inclusion of pairing effects. The thermal evolution of 1.4 M$_\odot$ stars from all parameter sets falls within the dark band. The thermal evolutions of the maximum mass stars are indicated by individual curves. \label{cool_SF}}
\end{figure}

Fig.~\ref{cool_SF} shows the thermal evolution of neutron stars in which pairing is present, as discussed above. In this case, the cooling of neutron stars with $1.4$ M$_\odot$ is very similar for all parameter sets and falls within the dark band in the figure. For neutron stars of higher masses we have two different scenarios, a ``faster" cooling for the higher symmetry energy parameter sets of the coupling schemes C1 and C2 and both parameter sets of C4, and a ``slower" cooling for the lower symmetry energy parameter sets of the coupling schemes C1 and C2 as well as both parameter sets of C3. These results are once more in agreement with the results shown in Fig.~\ref{Urca} if we simply consider that the thresholds for the DU process are effectively raised by the presence of pairing. 

In conclusion, the aim of this work was to update the non-linear realization of the hadronic sigma model to reflect new astrophysical observations and experimental nuclear data. More specifically, we looked for equations of state that reproduce small star radii (at the same time as large star masses) and a small symmetry energy slope at saturation. In addition, we verified that the thermal evolution of stars reproduced by our equations of state were in agreement with Cas A cooling data. The main features of these equations of state are provided in Table I and all of our results (in the format of data tables) can be provided upon request.

More specifically, we investigated the role of the vector-meson self-interaction terms within a chiral mean field formulation by performing a parameter scan. We also varied the coupling of the scalar-isovector meson to the nucleons. We found that the strength of the $\omega\rho$ term in the vector-meson self-interactions plays an important role in the equation of state, together with strange-vector-meson linear terms.

With respect to neutron star thermal evolution, our preliminary study showed that our results are in good agreement with modern cooling calculations. In particular, the schemes that include strong vector-meson self-interaction terms with low scalar-isovector coupling and the parametrizations that reproduce lower symmetry energies have an advantage over the others, as they exhibit a ``slower" cooling that can be identified for lower mass stars. This effect is further enhanced when pairing is included in the calculations, which was done by following the prescription used to model the thermal evolution of Cas A. In this case, the difference between ``slow" and ``fast" cooling stars was significantly enhanced.  A further, more in depth study of the effects of pairing in the thermal evolution of neutron stars using our equations of state will be reported in a future work. 

V. D. acknowledges the support from Dip. di Fisica e Scienze della Terra dell'Universita di Ferrara, INFN Sez. di Ferrara, Ferrara, and Helmholtz International Center for FAIR. R.N. acknowledges support from CAPES and CNPq. The authors acknowledge support from ÒNewCompStarÓ, COST Action MP1304.
\bibstyle{apsrev4-1}
\bibliography{apssamp}

\end{document}